\documentclass[reprint,amsmath,amssymb]{revtex4-1}
\usepackage[utf8x]{inputenc}
\usepackage{graphicx}
\usepackage{dcolumn}
\usepackage{bm}
\usepackage{amsthm}

\begin{document}

\preprint{APS/123-QED}

\title{dc conductivity as a geometric phase}
\author{Bal\'azs Het\'enyi}
\affiliation{Department of Physics \\
Bilkent University \\
TR-06800 Bilkent, Ankara, Turkey
}

\date{\today}

\begin{abstract}
The zero frequency conductivity ($D_c$), the criterion to distinguish between
conductors and insulators is expressed in terms of a geometric phase.  $D_c$
is also expressed using the formalism of the modern theory of polarization.
The tenet of Kohn [{\it Phys. Rev.} {\bf 133} A171 (1964)], namely, that
insulation is due to localization in the many-body space, is refined as
follows.  Wavefunctions which are eigenfunctions of the total current operator
give rise to a finite $D_c$ and are therefore metallic.  These states are also
delocalized.  Based on the value of $D_c$ it is also possible to distinguish
purely metallic states from states in which the metallic and insulating phases
coexist.  Several examples which corroborate the results are presented, as
well as a numerical implementation.  The formalism is also applied to the Hall
conductance, and the quantization condition for zero Hall conductance is
derived to be $\frac{e\Phi_B}{N h c} = \frac{Q}{M}$, with $Q$ and $M$
integers.
\end{abstract}


\maketitle

\section{Introduction} 

What makes conductors conducting and insulators insulating?  In classical
physics this question is answered by considering the localization of {\it
  individual} charge carriers.  Localized, bound charges do not contribute to
conduction.  Quantum mechanics has rendered the answering of this question
more difficult.  In band theory, conduction can be attributed to the density
of electron states at the Fermi level: if $\rho(\epsilon_F)\neq0$, the system
is conducting, if $\rho(\epsilon_F)=0$ it is insulating.  However, simple band
theory is not able to explain insulation of strongly correlated systems.  In
1964 Kohn suggested~\cite{Kohn64} that the criterion that distinguishes metals
from insulators is localization of the {\it total position} of all charge
carriers.  Kohn also derived~\cite{Kohn64} the quantum criterion of dc
conductivity, the Drude weight ($D_c$).

For several decades, testing Kohn's hypothesis was difficult, due to the fact
that in crystalline systems (systems with periodic boundary conditions) the
total position operator is ill-defined.  This limitation was overcome by the
modern theory of polarization~\cite{Fois88,Selloni87,Resta94,King-Smith93}, in
which the expectation value of the total position is cast in terms of a geometric
phase~\cite{Pancharatnam56,Berry84,Xiao10}.  The geometric phase arises upon
varying the crystal momentum across the Brillouin zone.  In numerical
applications the polarization is easiest to calculate in terms of the ground
state expectation value of the total momentum shift
operator~\cite{Resta98,Resta99}.  These developments have simplified the
calculation of the polarization considerably, and are now in widespread use in
electronic structure calculations.

Moulopoulos and Ashcroft~\cite{Moulopoulos92} have also suggested a connection
between conduction and a Berry phase related to the center of mass.  Recently,
the author has shown~\cite{Hetenyi12b} that the total current can be expressed
as a phase associated with moving the total position across the periodic cell,
and that it can be written as a ground state expectation value of the total
position shift operator.  We note that topological invariants can also
characterize metals~\cite{Zhou13} as well as insulators.

\section{Purpose} 

We demonstrate that $D_c$ can also be expressed in terms of a geometric phase.
The formal expression for $D_c$ derived here consists of an expectation value
of single-body operators and a geometric phase arising from the variation of
the total momentum {\it and} the total position.  Its form is similar to that
of the Hall conductance~\cite{Thouless82}.  The second term is also expressed
in terms of the total momentum and total position shift operators, in other
words, based on a formalism similar to that of the ``modern'' theory of
polarization.  The resulting formula establishes the precise connection
between localization and conductivity as suggested by Kohn~\cite{Kohn64}.  If
the ground state wavefunction of a system is an eigenstate of the total
current operator, $D_c$ is finite.  Such wavefunctions are also delocalized
according to the criterion defined by Resta~\cite{Resta98,Resta99}.  The
calculation of the Drude weight is also straightforward: for metals the $D_c =
\frac{\pi \alpha}{L}$.  (Eq. (\ref{eqn:alpha}), where $L$ denotes the size of
the system), for insulators it is zero.  For wavefunctions corresponding to
coexistence between metallic and insulating phases it holds that $0 < D_c <
\frac{\pi \alpha}{L}$.  One calculates the spread in total current, and if
this spread is zero, then $D_c = \frac{\pi \alpha}{L}$.  These results are
indepedent of dimensionality.  The formalism is also used to derive the Hall
conductance~\cite{Thouless82}, and a quantization condition for that quantity
being zero is derived.  The condition coincides with the well-known
experimental results for the fractional quantum Hall effect~\cite{Stormer99}.

\section{Definitions}  

Let $|\Psi\rangle$ denote the ground state wavefunction of an $N$ particle
system.  In coordinate space one can write $\Psi(x_1+X,...,x_N+X)$ where $X$
denotes a shift of all coordinates, or equivalently one can write in momentum
space $\Psi(k_1+K,...,k_N+K)$.  A wavefunction can be labeled by $X$ or $K$
($|\Psi(X)\rangle$, $|\Psi(K)\rangle$).  One can define the shift operators in
position or momentum space as
\begin{eqnarray}
\label{eqn:shift}
e^{-i\Delta K \hat{X}} |\Psi(K)\rangle &=& |\Psi(K+\Delta K)\rangle \\
e^{-i\Delta X \hat{K}} |\Psi(X)\rangle &=& |\Psi(X+\Delta X)\rangle, \nonumber
\end{eqnarray}
where $\hat{X} = \sum_{i=1}^N \hat{x}_i$, and $\hat{K} = \sum_{i=1}^N
\hat{k}_i$.  In lattice models the current operator in momentum space takes
the form $\hat{K}= \sum_{i=1}^N \sin(\hat{k}_i)$.  The explicit construction
of the shift operators is given in Refs. \cite{Hetenyi09,Hetenyi12b}.

\section{Main results}

\subsection{Conductivity as a geometric phase} 

The Drude weight~\cite{Kohn64} is defined as
\begin{equation}
\label{eqn:DC1}
D_c = \frac{\pi}{L} \frac{\partial^2 E(0)}{ \partial \Phi^2},
\end{equation}
where $\Phi$ denotes a perturbing field, and the derivative is the adiabatic
derivative.  The second derivative with respect to $\Phi$ can be expressed as
\begin{equation}
\label{eqn:DC2}
\frac{\partial^2 E(0)}{\partial \Phi^2} = \alpha + \gamma,
\end{equation}
where
\begin{equation}
\label{eqn:DC3}
\alpha = i\sum_j^{N}\langle \Psi |[\hat{\partial}_{k_j},\hat{\partial}_{x_j}]|\Psi\rangle,
\end{equation}
and where
\begin{eqnarray}
\label{eqn:gammaI}
\gamma =
- \frac{i}{2\pi} \int_{-\pi/L}^{\pi/L}\int_0^L \mbox{d} K \mbox{d} X \hspace{4cm}&\\
(\langle \partial_K \Psi(K,X) | \partial_X \Psi(K,X) \rangle 
-\langle \partial_X \Psi(K,X) | \partial_K \Psi(K,X) \rangle).&
\nonumber
\end{eqnarray}
This expression is derived in Appendix B.  $\gamma$ has the form of an
integrated Berry curvature over a surface in the two-dimensional space $K-X$,
and can be converted into a geometric phase by application of the Stokes
theorem.  Note that the Drude weight is the sum of two terms, one proportional
to the sum of the commutators of each momentum and position, and a
``commutator'' of the variables related to the total position and total
momentum of the system.

\subsection{Analog of $D_c$ based on the modern theory of
  polarization}  

$D_c$, in particular the term $\gamma$, can also be expressed using total
momentum and total position shift operators.  For charge carriers with mass
one, the one-body term is
\begin{equation}
\label{eqn:alpha}
  \alpha = \left\{ \begin{array}{rl}                                                                                                        
  N & \mbox{for continuyous models,}
  \\                                                                                                              
  -\frac{\langle \Psi | \hat{T}(0) | \Psi \rangle}{2} & \mbox{for lattice systems}.
       \end{array} \right.                                                                                                                   
\end{equation}
the geometric phase term can be written as
\begin{widetext}
\begin{equation}
\label{eqn:gammaII}
\gamma = -\lim_{\Delta X,\Delta K \rightarrow 0} \frac{1}{\Delta X \Delta
  K} \left[
\mbox{Im}\ln \frac{
\langle \Psi | e^{i\Delta K \hat{X}} e^{i\Delta X \hat{K}}|\Psi\rangle 
}{\langle \Psi | e^{i\Delta X \hat{K}}|\Psi\rangle} +
\mbox{Im}\ln \frac{
\langle \Psi |  e^{i\Delta X \hat{K}} e^{-i\Delta K \hat{X}}|\Psi\rangle 
}{\langle \Psi | e^{i\Delta X \hat{K}}|\Psi\rangle} \right].
\end{equation}
\end{widetext}
This expression is derived in Appendix C.

\section{Interpretation}

The first term of $D_c$, proportional to $\alpha$, is an extensive quantity, a
sum over single-body operators.  For any non-trivial system it is expected to
be finite.  For an insulator the many-body term (proportional to $\gamma$)
must cancel the single-body term.

We consider a general wavefunction of the form $\Psi(x_1,...,x_N)$
corresponding to an unperturbed ground state.  Acting on this function with
the shift operators according to the first and second terms of $\gamma$
(Eq. (\ref{eqn:gammaII})), respectively, results in
\begin{widetext}
\begin{eqnarray}
\label{eqn:gamma_1}
e^{i\Delta K \hat{X}} e^{i\Delta X \hat{K}} \Psi(x_1,...,x_N) &=& 
e^{iN\Delta K \Delta X} e^{i\Delta K \sum_{i=1}^N x_i} \Psi(x_1+\Delta
X,...,x_N + \Delta X), \\
 e^{i\Delta X \hat{K}} e^{-i\Delta K \hat{X}} \Psi(x_1,...,x_N) &=&
e^{-i\Delta K \sum_{i=1}^N x_i} \Psi(x_1+\Delta X,...,x_N + \Delta X).  \nonumber
\end{eqnarray}
\end{widetext}
Evaluating the scalar products, one can then show that apart from the term
$e^{iN\Delta K \Delta X}$ in Eq. (\ref{eqn:gamma_1}) the two terms in
Eq. (\ref{eqn:gammaII}) are complex conjugates of each other.  The term
$e^{iN\Delta K \Delta X}$ gives a contribution of $-N$ to the conductivity
cancelling the single-body term.  When this derivation is valid the system is
insulating.  This derivation, of course, has limits of validity, for example,
if discontinuities are present in the momentum distribution~\cite{Hetenyi12a}.

If the function $|\Psi\rangle$ is an eigenfunction of the current operator,
then $\gamma$ is zero, hence the system is metallic.  To show this, one
considers that the eigenvalue of the current operator for an unperturbed
ground state is zero, which means that the total position shift operator will
have no effect at all.  In this case the two terms of $\gamma$ are complex
conjugates of each other, and their sum will have no imaginary part.

If a wavefunction is an eigenstate of the total current operator, it also
follows that the system is delocalized.  Indeed the localization criterion
defined by Resta~\cite{Resta98,Resta99} is
\begin{equation}
\label{eqn:sigX2}
\sigma_X^2 = -\frac{2}{\Delta K^2} \mbox{Re} \ln 
\langle \Psi | e^{-i\Delta K \hat{X}} | \Psi \rangle.
\end{equation}
The function resulting from the total momentum shift operator acting on an
eigenfunction of the total current will be orthogonal to the original
function, resulting in a divergent $\sigma_X^2$.  

To decide whether a particular ground state eigenfunction is an eigenfunction
of the current one can calculate the spread in current~\cite{Hetenyi12b},
defined as
\begin{equation}
\sigma_K^2 = -\frac{2}{\Delta X^2} \mbox{Re} \ln 
\langle \Psi | e^{-i\Delta X \hat{K}} | \Psi \rangle.
\end{equation}
If $\sigma_K$ is zero then the wavefunction is indeed a current eigenstate,
the system is metallic, moreover $\gamma=0$ and the $D_c= \frac{\pi
  \alpha}{L}$.  Otherwise the wavefunction corresponds to an insulating state.
To show this one can use the fact that for an eigenfunction of the current
with eigenvalue zero the expectation value $\langle \Psi | e^{-i\Delta X
  \hat{K}} | \Psi \rangle=1$, must give one, but for any other case $\langle
\Psi | e^{-i\Delta X \hat{K}} | \Psi \rangle<1$.  In calculating conductivity,
one can also use Eq. (\ref{eqn:sigX2}), but this quantity is expected to
diverge when the system becomes metallic, hence calculations based on
$\sigma_K$ can be expected to be more stable.

A wavefunction can also be a linear combination of an eigenstate
of the current operator and a localized state corresponding to the coexistence
of the insulating and metallic states.  In this case the single body
term will be partially cancelled by the many-body term and a finite Drude
weight will result, but in that case $D_c$ will be smaller than the
contribution due to single-particle operators (for continuous models $D_c<N$).

\section{Examples}

\subsection{Fermi sea, BCS}  

For both the Fermi sea and BCS wavefunctions $D_c=\frac{\pi \alpha}{L}$.  The
Fermi sea is diagonal in the momentum representation and corresponds to an
eigenstate of $\hat{K}$ with eigenvalue zero.  A BCS wavefunction consists of
a linear combination of wavefunctions with different number of particles, but
all have eigenvalue of $\hat{K}=0$, and the argument for the Fermi sea
extends.

\subsection{Gutzwiller metal}  

The Gutzwiller variational wavefunction was proposed to understand the Hubbard
model~\cite{Gutzwiller63,Hubbard63,Kanamori63}, and is of the form
\begin{equation}
|\Psi_G(\tilde{\gamma})\rangle = e^{-\tilde{\gamma}\sum_i 
\hat{n}_{i\uparrow}\hat{n}_{i\downarrow}
} | FS \rangle.
\end{equation}
The state $|FS\rangle$ denotes the Fermi sea, out of which doubly occupied
sites are projected out via the projector $e^{-\tilde{\gamma}\sum_i
  \hat{n}_{i\uparrow}\hat{n}_{i\downarrow}}$.  This wavefunction has been
shown~\cite{Millis91,Dzierzawa97} to be metallic for finite values of the
variational parameter $\tilde{\gamma}$, ($D_c=\frac{\alpha \pi}{L}$).

Indeed, the geometric phase term $\gamma$ vanishes.  To see this, consider
that the shift operator $e^{i\Delta X \hat{K}}$ commutes with the projector
$e^{-\tilde{\gamma}\sum_i \hat{n}_{i\uparrow}\hat{n}_{i\downarrow}}$, since
shifting the position of every particle will not affect the number of doubly
occupied sites~\cite{Hetenyi12b}.  Thus $e^{i\Delta X \hat{K}}$ will operate
on the Fermi sea, which has eigenvalue $\hat{K}|FS\rangle = 0$, and then the
same reasoning applies as in the case of the Fermi sea.

\subsection{Baeriswyl insulating wavefunction for a spinless system}  

An insulating variational solution for spinless fermions on a lattice with
nearest neighbor interaction ($t$-$V$ model) in one dimension, is the
Baeriswyl wavefunction~\cite{Valenzuela03}, which in this case has the form
\begin{equation}
|\Psi_B(\tilde{\alpha})\rangle = \prod_{\mbox{RBZ}} [
e^{-\tilde{\alpha} \epsilon_k} c^\dagger_k 
+
e^{\tilde{\alpha} \epsilon_k} c^\dagger_{k+\pi}
]|0\rangle,
\end{equation}
where the product is over the reduced Brillouin zone.  This wavefunction is
easily shown to be insulating~\cite{Valenzuela03}, hence we expect that it
gives $D_c=0$.

This can be shown readily by considering again the action of the shift
operators on $|\Psi_B(\tilde{\alpha})\rangle$.  The scalar products in $\gamma$
evaluate to
\begin{widetext}
\begin{eqnarray}
\langle \Psi_B(\tilde{\alpha})|e^{i\Delta K \hat{X}} e^{i\Delta X \hat{K}} |\Psi_B(\tilde{\alpha})\rangle
 &=&
\prod_{\mbox{RBZ}} [
e^{i\Delta X \sin(k+\Delta K)}e^{-\tilde{\alpha} (\epsilon_k+\epsilon_{k+\Delta K})} 
+
e^{-i\Delta X \sin(k+\Delta K)} e^{\tilde{\alpha} (\epsilon_k + \epsilon_{k+\Delta
    K})} ], \\
\langle \Psi_B(\tilde{\alpha})| e^{i\Delta X \hat{K}} e^{-i\Delta K \hat{X}}|\Psi_B(\tilde{\alpha})\rangle
 &=&
\prod_{\mbox{RBZ}} [
e^{i\Delta X \sin(k)}e^{-\tilde{\alpha} (\epsilon_k+\epsilon_{k-\Delta K})} 
+
e^{-i\Delta X \sin(k)} e^{\tilde{\alpha} (\epsilon_k + \epsilon_{k-\Delta
    K})} ].
\nonumber
\end{eqnarray}
\end{widetext}
Substituting into the definiton of $\gamma$ and taking the limits $\Delta
K,\Delta X \rightarrow 0$ lead to $D_c=0$ as expected for an insulating state.
The above derivation is also valid for the mean-field spin or charge-density
wave solutions of strongly correlated lattice models.

\subsection{Anderson localized system} 

We have evaluated the above formula for a model which exhibits Anderson
localization~\cite{Anderson58}, with Hamiltonian of the form
\begin{equation}
H = -t \sum_i c^\dagger_i c_{i+1} + \mbox{H. c.} + U \sum_i \xi_i n_i,
\end{equation}
where $\xi_i$ is a number drawn from a uniform Gaussian distribution.  By
diagonalizing the Hamiltonian we have calculated the localization
parameter~\cite{Resta98,Resta99} for different system sizes, and have found
that the larger system sizes are always more localized for finite $U$ (results
not shown).  We have also calculated the Drude weight and the quantity
$\sigma_K$.  The results are shown in Table I.
\begin{table}
\begin{tabular}{|c||c|c|c|c|}
\hline
$U$ & $\sigma_K$ & $D_c\times L /\pi$ & $-\frac{\langle T \rangle}{2}$ & $\sigma_X$  \\ \hline
$0$ & $0$ & $327.95$ & $327.95$ & ---  \\ 
$1$ & $5.8(2)$ & $0.01154(4)$ & $297(7)$ & $38(4)$     \\ 
$2$ & $9.8(2)$ &$0.0087(1)$ & $233(3)$ & $17.8(9)$     \\ 
$3$ & $12.8(2)$ &$0.0066(2)$ & $175(5)$ & $11.7(5)$     \\ 
$4$ & $15.1(2)$ &$0.0051(2)$ & $136(5)$&$8.4(4)$     \\ 
$5$ & $16.7(3)$ &$0.0041(2)$ &$110(5)$ &$6.5(3)$     \\ 
\hline
\end{tabular}
\caption{Results from diagonalization of Anderson localization model for a
  system with $1024$ lattice sites and $512$ particles.  $\Delta K = \Delta X
  = 0.001$ }
\label{tab:Anderson}
\end{table}           

For the metallic state $\sigma_K$ gives zero as expected, and the Drude weight
is equal to minus one-half the kinetic energy.  For all insulating cases the
Drude weight is very near zero, in particular if one compares its magnitude to
that of the kinetic energy.  While one can calculate the Drude weight
directly, this may be difficult in some applications, since phases have to be
evaluated.  However evaluating the kinetic energy and the spread in current
allows the determination of the Drude weight unambiguously.

\section{Hall conductance}  

The Hall conductance can also be expressed in terms of a Berry
phase~\cite{Thouless82}, similar in form to the conductivity derived above
(Eq. (\ref{eqn:gammaI})).  It is possible to express the Hall conductance as a
ground state observable.~\cite{Neupert12a,Neupert12b} Here we express it via
shift operators, and derive a quantization condition for zero Hall conductance
in a quantum Hall system.  The momentum shift operators in this case take
forms which are different from those used in expressing dc conductivity.

Our starting point is the form derived by Thouless {\it et
  al.}~\cite{Thouless82},
\begin{equation}
\sigma^H_{xy} = \frac{ie^2}{2 \pi h} \int \mbox{d}K_x \mbox{d}K_y
[\langle \partial_{K_x} \Psi |\partial_{K_y} \Psi \rangle - \mbox{H.c.}],
\end{equation}
which, using the formalism above converts to
\begin{widetext}
\begin{equation}
\sigma^H_{xy} = \frac{e^2}{h} \lim_{\Delta K_x \Delta K_y \rightarrow 0} \frac{1}{\Delta K_x \Delta K_y} 
\left[ \mbox{Im}\ln \frac{
\langle \Psi | U_x(\Delta K_x) U_y(\Delta K_y) |\Psi\rangle 
}{\langle \Psi | U_y(\Delta K_y) |\Psi\rangle} +
\mbox{Im}\ln \frac{
\langle \Psi |  U_y(\Delta K_y) U_x(-\Delta K_x)|\Psi\rangle 
}{\langle \Psi | U_y(\Delta K_y) |\Psi\rangle} \right],
\label{eqn:Hall}
\end{equation}
\end{widetext}
where $U_x(\Delta K_x)$ and $U_y(\Delta K_y)$ are momentum shift operators in
the $x$ and $y$ directions. Using the forms of the total momentum shift
operators in Eqs. (\ref{eqn:shift}) (applicable when the wavefunctions can be
written in the coordinate or momentum representations) we can show that in the
limit $\Delta K_x,\Delta K_y \rightarrow 0$ the Hall conductivity takes the
form
\begin{equation}
\sigma^H_{xy} = \frac{ie^2}{h} \sum_i\langle \Psi |[\hat{x}_i,\hat{y}_i]|\Psi\rangle.
\end{equation}

Using Eq. (\ref{eqn:Hall}) applied to a Landau state one can also derive a
quantization condition for the values of the magnetic field at which
$\sigma^H_{xy}$ must be zero.  A Landau level has the form 
\begin{equation}
\psi(x,y) = e^{ik_x x} \phi_n(y-y_0),
\end{equation}
where $y_0=k_x\frac{\hbar c}{eB}$.  As far as the $x$ direction is concerned
this function is neither in the momentum nor in the position representations.
However, the momentum shift operators can be constructed, considering that a
momentum shift in the $x$-direction is also a position shift in the $y$
direction.  It is easy to check that in this case
\begin{equation}
U_x(\Delta K_x) = e^{i\Delta K_x x} e^{i \Delta Y k_y},
\end{equation}
with $\Delta Y = \Delta K_x \frac{\hbar c}{eB}$.  The momentum shift in the
$y$ direction remains
\begin{equation}
U_y(\Delta K_y) = e^{i\Delta K_y y}.
\end{equation}
Applying the shift operators to the Landau state results in
\begin{widetext}
\begin{eqnarray}
\label{eqn:shifts_Hall}
U_x(\Delta K_x) U_y(\Delta K_y) \psi(x,y) &=& e^{i\Delta K_y (y-y_0)} e^{i\Delta
  K_x x} \psi(x,y+\Delta y), \\  \nonumber
U_y(\Delta K_y) U_x(-\Delta K_x) \psi(x,y) &=&
e^{i\Delta K_y \Delta y} e^{i\Delta K_y (y-y_0)} e^{i\Delta K_x x}
\psi(x,y-\Delta y), 
\end{eqnarray}
\end{widetext}
where $\Delta y = \Delta K_x \frac{\hbar c}{eB}$.  If $\Delta K_x \Delta y =
\Delta K_x \Delta K_y \frac{\hbar c}{eB}=2\pi M$, with $M$ integer, then the
phase in the second of Eqs. (\ref{eqn:shifts_Hall}) is one, and in this case
taking the limits $\Delta K_x, \Delta K_y \rightarrow 0$ results in a Hall
conductance of zero.  We tcan ake the momentum shifts to be $\Delta K_x = q_x
\frac{2\pi}{L_x}$ and $\Delta K_y = q_y \frac{2\pi}{L_x}$, with $q_x,q_y$
integers, which corresponds to equivalent states for the adiabatic
case~\cite{Laughlin81,Byers61} it follows that for a system with $N$ particles
the quantization condition is
\begin{equation}
\label{eqn:phi}
\frac{e\Phi_B}{N h c} = \frac{Q}{M},
\end{equation}
where $\Phi_B$ denotes the magnetic flux, and $Q$ is an integer.  Indeed, the
maxima in the Hall resistivity occur~\cite{Stormer99} precisely at
values of the magnetic flux given by Eq. (\ref{eqn:phi}).

\section{Conclusion}  In this work it was shown that the zero frequency
conductivity can be expressed in terms of a Berry phase.  Subsequently the
conductivity was also expressed in terms of shift operators (total momentum
and total position) leading to expressions which provide clear physical
insight, as well as a good starting point for numerical work.  It was argued
that a metallic state is one which is the eigenstate of the total current
operator.  Such states were also shown to be delocalized.  In this case the dc
conductivity takes its maximum possible value for a given system (proportional
to the number of charge carriers for continuous models).  These conclusions
were supported by analytic and numerical calculations on a number of examples,
both metallic and insulating.  If the wavefunction is a linear combination of
a total current eigenstate and an insulating wavefunction then a finite dc
condutivity results which is smaller than the allowed maximum.  Hence, based
on the value of the dc conductivity it is possible to distinguish metallic and
insulating states from ones in which conducting and insulating states coexist.
Subsequently the formalism was used to express the Hall conductance, and to
derive the quantization condition at which the Hall conductance is zero.  The
condition coincides with the well-known experimental results.

\section*{Acknowledgments}
The author acknowledges a grant from the Turkish agency for basic research
(T\"UBITAK, grant no. 112T176).

\section*{Appendices}

\subsection*{APPENDIX A: Perturbed Hamiltonian}

The dc conductivity~\cite{Kohn64} is proportional to the second derivative of
the ground state energy with respect to the Peierls phase $\Phi$ at $\Phi=0$.
For a continuous system, taking the mass of charge carriers to be unity, the
Hamiltonian has the form
\begin{equation}
\label{eqn:HPhi}
\hat{H}(\Phi) = \sum_j \frac{(\hat{k}_j + \Phi)^2}{2} + \hat{V},
\end{equation}
in the case of discrete models one can write
\begin{equation}
\hat{H}(\Phi) = \hat{T} + \hat{V},
\end{equation}
with
\begin{equation}
\hat{T}(\Phi) = - \sum_j t e^{i\Phi} c_{j+1}^\dagger c_j + \mbox{H. c.}.
\end{equation}
For a detailed discussion see Refs. \cite{Kohn64} and \cite{Essler05})
For both continuous and lattice Hamiltonians it holds that
\begin{equation}
H'(0) = i[\hat{H},\hat{X}] = \hat{K},
\end{equation}
and
\begin{equation}
H''(0) = i[\hat{K},\hat{X}],
\end{equation}
where $\hat{X}$($\hat{K}$) are defined as
\begin{eqnarray}
\hat{X} =& \sum_j \hat{x}_j \\ \nonumber
\hat{K} =& \sum_j \hat{k}_j,
\end{eqnarray}
for continuous systems and 
\begin{eqnarray}
\hat{X} =& \sum_j j \hat{n}_j \hspace{2cm}\\ \nonumber
\hat{K} =& -it \sum_j c_{j+1}^\dagger c_j + H.c.,
\end{eqnarray}
for lattice models.  One can also write $H''(0)$ as a sum of one-body
operators as
\begin{equation}
\label{eqn:com_1body}
H''(0) = -\sum_j[\hat{k}_j,\hat{\partial}_{k_j}]= -\sum_j[\hat{\partial}_{x_j},\hat{x}_j].
\end{equation}
One can also show that
\begin{equation}
  H''(0) = \left\{ \begin{array}{rl}                                                                                                        
  N & \mbox{for continuous models,}
  \\                                                                                                              
  -\hat{T}(0) & \mbox{for lattice systems}.
       \end{array} \right.                                                                                                                   
\end{equation}
One can expand the Hamiltonian and the ground state wavefunction up to second
order as
\begin{eqnarray}
H(\Phi) \approx& H(0) + \Phi H'(0) + \frac{\Phi^2}{2} H''(0) \\
|\Psi(\Phi)\rangle \approx& |\Psi(0)\rangle + \Phi |\Psi'(0)\rangle +
\frac{\Phi^2}{2} |\Psi''(0)\rangle \nonumber
\end{eqnarray}
and express the second derivative of the ground state energy with respect to
$\Phi$ at $\Phi=0$ as
\begin{eqnarray}
\partial^2_\Phi E(\Phi)|_{\Phi=0} = \langle \Phi(0)| H''(0) |\Phi(0) \rangle \hspace{3cm}\\
+ 2 \langle \Phi'(0)| H'(0) |\Phi(0) \rangle 
+ 2 \langle \Phi(0)| H'(0) |\Phi'(0) \rangle.
\nonumber
\end{eqnarray}

\subsection*{APPENDIX B: dc conductivity as a geometric phase}

In this appendix the dc conductivity is derived in terms of a geometric
phase.  As shown in Ref. \cite{Hetenyi12b} the first derivative of the ground
state energy with respect to $\Phi$ for a continuous Hamiltonian is given by
\begin{equation}
\partial_\Phi E(\Phi) = \alpha \Phi - 
 \frac{i}{L} \int_0^L \langle \Psi(X;\Phi ) | \partial_X | \Psi(X;\Phi)\rangle,
\end{equation}
where
\begin{equation}
\label{eqn:alpha_app}
  \alpha = \left\{ \begin{array}{rl}                                                                                                        
  N& \mbox{for continuous models,}
  \\                                                                                                              
  -\langle \Psi | \hat{T}(0) | \Psi \rangle & \mbox{for lattice systems}.
       \end{array} \right.                                                                                                                   
\end{equation}
Taking the derivative with respect to $\Phi$ and setting $\Phi$ to zero
results in
\begin{eqnarray}
\partial^2_\Phi E(\Phi)|_{\Phi=0} = \alpha - 
 \frac{i}{L} \int_0^L \mbox{d}X\hspace{4cm}\\ \nonumber
[\langle \partial_\Phi \Psi(X ) | \partial_X | \Psi(X)\rangle
+
\langle \Psi(X ) | \partial_X | \partial_\Phi \Psi(X)\rangle].
\end{eqnarray}
Since $\Phi$ corresponds to a shift in the crystal momentum $K$ the derivative
with respect to $\Phi$ can be replaced with a derivative with respect to $K$.
Subsequently an average over $K$ can be taken, resulting in
\begin{equation}
\label{eqn:DC_phase}
\partial^2_\Phi E(\Phi)|_{\Phi=0} = \alpha + \gamma,
\end{equation}
with
\begin{eqnarray}
\gamma = - 
 \frac{i}{2\pi} \int_0^L \int_{-\pi/L}^{\pi/L} \mbox{d}X\mbox{d}K \hspace{2cm}\\ \nonumber
[\langle \partial_K \Psi(X,K) | \partial_X  \Psi(X,K)\rangle
-
\langle \partial_X \Psi(X,K) | \partial_K \Psi(X,K)\rangle].
\end{eqnarray}
The quantity $\gamma$ in Eq. (\ref{eqn:DC_phase}) is a surface integral over a Berry
curvature, which can be converted into a line integral around the included
surface via Stokes theorem, as for the Hall conductivity~\cite{Thouless82}.

The quantity $\alpha$ can be written with the help of
Eq. (\ref{eqn:com_1body}) as
\begin{equation}
\alpha = i\sum_j \langle \Psi | [\partial_{x_j},\partial_{k_j}] | \Psi
\rangle.
\end{equation}
In other words the conductivity corresponds to the difference between the sum
of one body commutators of the position and momenta and the commutator of the
total position and total momentum.

\subsection*{APPENDIX C: dc conductivity in terms of shift operators}

Our starting point is the current~\cite{Hetenyi12b} written in terms of shift
operators~\cite{Hetenyi09},
\begin{equation}
\partial_\Phi E(\Phi) = \alpha \Phi - 
\frac{1}{\Delta X}\mbox{Im}\ln \langle \Psi (\Phi)| e^{i\Delta X \hat{K}} |\Psi (\Phi)\rangle.
\end{equation}
Taking the derivative with respect to $\Phi$ results in
\begin{equation}
\partial_\Phi E(\Phi) = \alpha + \gamma,
\end{equation}
with
\begin{widetext}
\begin{equation}
\label{eqn:gamma1}
\gamma = \frac{1}{\Delta X} \mbox{Im} 
\left[
\frac{\langle \partial_\Phi \Psi(\Phi) |e^{i\Delta X \hat{K}}| \Psi(\Phi)
  \rangle}{\langle \Psi(\Phi) |e^{i\Delta X \hat{K}}| \Psi(\Phi) \rangle}
+
\frac{\langle  \Psi(\Phi) |e^{i\Delta X \hat{K}}| \partial_\Phi\Psi(\Phi)
  \rangle}{\langle \Psi(\Phi) |e^{i\Delta X \hat{K}}| \Psi(\Phi) \rangle}
\right]
\end{equation}
\end{widetext}
We can set the derivative in $\Phi$ equal to the derivative in the crystal
momentum, and set $\Phi=0$.  For now we will consider only the first term in
Eq. (\ref{eqn:gamma1}) but the steps for the second term are essentially
identical.  We can write this term as
\begin{equation}
\frac{1}{\Delta X \Delta K} \mbox{Im} 
\left[
\frac{\Delta K \langle \partial_K \Psi(0) |e^{i\Delta X \hat{K}}| \Psi(0)
  \rangle}{\langle \Psi(0) |e^{i\Delta X \hat{K}}| \Psi(0) \rangle}
\right],
\end{equation}
where we have divided and multiplied by $\Delta K$.  For small $\Delta K$ we
can replace this term with
\begin{equation}
\frac{1}{\Delta X \Delta K} \mbox{Im} \ln
\left[ 1 + 
\frac{\Delta K \langle \partial_K \Psi(0) |e^{i\Delta X \hat{K}}| \Psi(0)
  \rangle}{\langle \Psi(0) |e^{i\Delta X \hat{K}}| \Psi(0) \rangle}
\right],
\end{equation}
which can be converted to
\begin{equation}
\frac{1}{\Delta X \Delta K} \mbox{Im} \ln
\left[
\frac{\langle \Psi(\Delta K) |e^{i\Delta X \hat{K}}| \Psi(0)
  \rangle}{\langle \Psi(0) |e^{i\Delta X \hat{K}}| \Psi(0) \rangle}
\right],
\end{equation}
and using the total momentum shift operator results in
\begin{equation}
\frac{1}{\Delta X \Delta K} \mbox{Im} \ln
\left[
\frac{\langle \Psi |e^{i\Delta K \hat{X}} e^{i\Delta X \hat{K}}| \Psi
  \rangle}{\langle \Psi |e^{i\Delta X \hat{K}}| \Psi \rangle}
\right].
\end{equation}
Applying exactly the same steps to the second term of Eq. (\ref{eqn:gamma1})
results in
\begin{widetext}
\begin{equation}
\label{eqn:gamma2}
\gamma = \frac{1}{\Delta X\Delta K} 
\left[
\mbox{Im} \ln
\left(
\frac{\langle \Psi |e^{i\Delta K \hat{X}} e^{i\Delta X \hat{K}}| \Psi
  \rangle}{\langle \Psi |e^{i\Delta X \hat{K}}| \Psi \rangle}
\right) + 
\mbox{Im} \ln
\left(
\frac{\langle \Psi | e^{i\Delta X \hat{K}} e^{-i\Delta K \hat{X}}| \Psi
  \rangle}{\langle \Psi |e^{i\Delta X \hat{K}}| \Psi \rangle}
\right)
\right],
\end{equation}
\end{widetext}
which is the discretized form for the Drude weight.


\begin{thebibliography}{9}

\bibitem{Kohn64} W. Kohn, {\it Phys. Rev.} {\bf 133} A171 (1964).

\bibitem{Fois88} E.~S. Fois, A. Selloni, M. Parrinello and R. Car, {\it
    J. Phys. Chem.}, {\bf 92} 3268 (1988).

\bibitem{Selloni87} A. Selloni, P. Carnevali, R. Car and M. Parrinello, {\it
    Phys. Rev. Lett.}, {\bf 59} 823 (1987).

\bibitem{Resta94} R. Resta, {\it Rev. Mod. Phys.} {\bf 66} 899 (1994).

\bibitem{King-Smith93} R. D. King-Smith and D. Vanderbilt, {\it Phys. Rev. B} {\bf 47} 1651 (1993).

\bibitem{Pancharatnam56} S. Pancharatnam, {\it Proc. Indian Acad. Sci. A} {\bf
  44} 247 (1956).

\bibitem{Berry84} M. V. Berry, {\it Proc. Roy. Soc. London} {\bf A392} 45
  (1984).

\bibitem{Xiao10} D. Xiao, M.-C. Chang, and Q. Niu, {\it Rev. Mod. Phys.} {\bf
  82} 1959 (2010).
\bibitem{Resta98} R. Resta, {\it Phys. Rev. Lett.} {\bf 80} 1800 (1998).

\bibitem{Resta99} R. Resta and S. Sorella, {\it Phys. Rev. Lett.} {\bf 82}
   370 (1999).

\bibitem{Moulopoulos92} K. Moulopoulos and N. W. Ashcroft, {\it Phys. Rev. B}, {\bf 45}  11518 (1992).

\bibitem{Hetenyi12b} B. Het\'enyi, {\it J.  Phys. Soc. Japan} {\bf 81 } 124711
  (2012).

\bibitem{Zhou13} J.-H. Zhou, H. Jiang, Q. Niu, and J.-R. Shi, {\it Chinese
  Phys. Lett.} {\bf 30} 027101 (2013).

\bibitem{Thouless82} D. J. Thouless, M. Kohmoto, M. P. Nightingale, and M. den Nijs
 {\it Phys. Rev. Lett} {\bf 49} 405 (1982).

\bibitem{Stormer99} H.~L. St\"ormer, D.~C. Tsui, and A.~C. Gossard, ,  {\it
 Rev. Mod Phys.}, {\bf 71} S298 (1999).

\bibitem{Hetenyi09} B. Het\'enyi, {\it J. Phys. A}, {\bf 42}  412003 (2009).


\bibitem{Hetenyi12a} B. Het\'enyi, {\it J. Phys. Soc. Japan} {\bf 81} 023701
  (2012).

\bibitem{Gutzwiller63} M.~C. Gutzwiller, {\it Phys. Rev. Lett.}, {\bf 10} 159 (1963).
  

\bibitem{Hubbard63} J. Hubbard, {\it Proc. Roy. Soc.} {\bf A276} 238 (1963).
  

\bibitem{Kanamori63} J. Kanamori, {\it Prog. Theoret. Phys.} {\bf 30} 275 (1963).

\bibitem{Millis91} A. J. Millis and S. N. Coppersmith, {\it Phys. Rev. B} {\bf
  43} 13770 (1991).

\bibitem{Dzierzawa97} M. Dzierzawa, D. Baeriswyl, and L.~M. Martelo, {\it
  Helv. Phys. Acta}, {\bf 70} 124 (1997).

\bibitem{Valenzuela03} B. Valenzuela, S. Fratini, and D. Baeriswyl,  {\it
 Phys. Rev. B}, {\bf 68} 045112 (2003).

\bibitem{Anderson58} P.~W. Anderson,  {\it
 Phys. Rev.}, {\bf 109} 1492 (1958).

\bibitem{Neupert12a} T. Neupert, L. Santos, C. Chamon, and C. Mudry,  {\it
 Phys. Rev. B}, {\bf 86} 165133 (2012).

\bibitem{Neupert12b} T. Neupert, L. Santos, S. Ryu, C. Chamon, and C. Mudry,  {\it
 Phys. Rev. B}, {\bf 86} 035125 (2012).

\bibitem{Laughlin81} R.~B. Laughlin,  {\it Phys. Rev. B}, {\bf 23} 5632 (1981).

\bibitem{Byers61} N. Byers and C.~N. Yang: Phys. Rev. Lett.
  {\bf 7} (1961) 46.

\bibitem{Essler05} F.~H.~L. Essler, H. Frahm, F. G\"ohmann, A. Kl\"umper,
  and V.~E. Korepin, {\it The One-Dimensional Hubbard Model}, Cambridge
  University Press, (2005).

\end{thebibliography}
\end{document}